\documentstyle[tighten,12pt,russian,aps]{revtex}

\begin{document}
\author{Yu.L.Klimontovich}
\address{1179899 Rusiian Moscow\\
Moscow M.V.Lomonosov State University\\
Physical Department}
\title{To Phenomenological Theory of Superfluidity.\\
Superfluidity - Viscousless Fflow in Viscous Medium}
\maketitle

\begin{abstract}
Despite of remarkable successes the theory of superfluidity, until now we
have no a physical explanation of an opportunity of existence of viscousless
flow of liquid helium in viscous medium. The existence of superfluid flow
becomes possible due to occurence of flicker noise and appropriate residual
temporary correlations of the superfluid velocity fluctuations.
\end{abstract}

\section{Introduction}

In 1932 Keesom and Klausius (see Keesom, 1949) have found out in the region
temperatures $T_{C}$ = 2.19 $^{0}$ a anomalous temperature dependence of the
heat capacity. That to emphasize distinction of states of the helium higher
and below of the of the phase transition point, the names were entered: ''
helium I '' for temperatures $T>T_{C}$ and '' helium II '' for temperatures $%
T<T_{C}.$

In 1938 (Kapitza, 1938, 1941, 1944) has found out, that the helium has
superfluidity - the ability of vicousless of flow through thin cracks and
capillaries. He has found out also presence in capillaries the opposite flow
of viscous - ''normal'' and viscousless - ''superfluid'' component liquid
helium . 

In the Kapitza experiments the viscous flow run out from the small
vessel is filled by liquid helium. The flow is caused by the warming up
helium in the vessel and it is discovered by a deviation of a target. It is
essential, that thus the level of helium in he vessel remain constant. This
indicates on the presence of a opposite flow - the flow of the superfluid
helium.

To explain this phenomenon he accepts '' two-liquid model '' of helium II,
offered in work (Tisza, 1938) and essential developed by (Landau, 1941), but
in (Kapitza, 1944) he writes:

'' If this theoretical supposition was not so full supported by the
experimental proofs, it would sound as idea, which it is very difficult to
recognize reasonable. ''

Thus, two-liquid model, even for the physicist of so high level, it is not
represented rather clear.

Use of two-liquid model is justified by that on its basis it is possible to
describe observable phenomena. In particulars, it gives an explanation the
existence of the second sound (Landau, 1944), explains results of the
Andronikashvily experiences with rotating helium (Andronokashvily, 1940).

Despite of these doubtless successes, the physical picture of the phenomenon
superfluidity remains nevertheless not quite clear. Two basic question
remain without the answer:

1. Physical nature of phase transition helium I - helium II.

2. Physical explanation of an opportunity of existence of viscousless flow
in viscous medium.

The basic purpose of the present work - to discuss the possible answers on
these questions on the following suppositions:

1). Superfluidity the macroscopic phenomenon in a dissipative continuous
medium is. The description of superfluidity will be curried out on the basis
of the corresponding equations of the mechanics of continuous medium.

2). Helium II is the example of a quantum liquid, as length of de Broglie
wave is order of the average distance between atoms. However, in the
approximation of continuous medium the de Broglie length is much less of
size of \ a ''point'', \ in which many particles contain. Due to this
condition the description of superfluidity on a basis of the
phenomenological kinetic equations for classical distribution function it is
possible.

3). The superfluidity by two phenomena is caused .

First, by the phase transition of the second order, as a result of which
arise not only fast, but also slow (coherence) relaxation processes. The
last provide a spatial coherence on scales cracks or capillaries.

Secondly, there is a reorganization of structure of viscous friction. It is
caused by occurrence of flicker noise (''$1/f$-noise) of hydrodynamic
velocity fluctuations. The appropriate distribution on wave numbers
represents an analogue of \ the ''bose-condensation'', as the dispersion of
wave numbers become proportional to frequency for $\omega \rightarrow 0$.

\section{Phase transition helium I - helium II}

A level with the phenomenological theory, the microscopic theory of
superfluidity is developed also. The basic work in this direction the
Bogolubov paper ''To the theory of superfluidity'' is (Bogolubov, 1947). By
the object of research in the Bogolubov paper a weak nonideal Bose-gas was
served.

\subsection{The condensation of a weakly nonideal gas}

In the classic Einstein (1924, 1925) papers is shown, that the continuous
Bose-Einstein distribution it is valid only at temperatures $T>3.14^{0}K$.
In the appropriate critical point the distribution of particles on momenta
breaks up to two parts: continuous distribution with zero value of the
chemical potential and the ''condensate'' with the distribution $\delta (p)$%
. On a measure downturn of temperature the number of particles with a zero
momentum grows and at zero temperature coincides with full number of
particles. At the critical temperature the number of particles in a
condensate is equal to zero.

Attempts (Tisza, 1938 and London 1954) to explain superfluidity of helium on
the basis of the Bose-condensation phenomenon of ideal gas were not
successful.

\subsection{Weak nonideal Bose-gas. Bogolubov theory}

In the ideal gas the condensate does not form the connected collective and
therefore has not property of superfluidity.

In a weak nonideal gas because of interaction of particles, as Bogolubov has
shown , the condensate forms collective and at it the motion as the whole
only at the birth of collective elementary excitation is possible. At small
momenta the energy of excitations is defined by classical expression:

\begin{equation}
\varepsilon (p)=v_{S}p,\quad v_{S}=\sqrt{n\frac{\nu (|p=0|)}{m}}p=\sqrt{%
\frac{4\pi \hbar ^{2}a}{m^{2}}n}.  \label{S.1.10}
\end{equation}
which answers to a phonon part of the Landau spectrum. The speed of sound $%
v_{S}$ via the amplitude $a$ of the Born scattering is expressed

The Bogolubov spectra is valid that the pushing away of atoms prevails above
their attraction ones. For the liquid helium the quantity $a$ is significant
less of the mean distance $r_{av}$ between atoms, therefore in the theory of
superfluidity it is possible introduce the corresponding small density
parameter $\varepsilon _{S}=na^{3}$.

At zero temperature the number of particles of nonideal gas in the
condensate is less than the complete number of particles. The difference is
defined by the small parameter $\varepsilon _{S}$. Thus for the weak
nonideal gas the basic part of atoms of gas has a zero momentum.

The next years the essential development of the microscopic theory of weakly
nonideal Bose-gas at $T=0$ has undergone. Among many it is necessary to
allocate the followings.

In the paper (Bogolubov, Zubarev, 1955) for Bose-gas the method of
collective variables was used. In the paper (Belyaev, 1958) the perturbation
theory on small density parameter $\varepsilon _{S}=na^{3}$ was developed$.$
In the series of papers (Kirpatrick and Dorfman, 1985) the kinetic
description of collective behavior of Bose-gas was carried out.

At last, in the papers (Tserkovnikov, 1992, 1995) the most detailed
description of weakly nonideal Bose-gas at $T=0$ is given and the
comparative analysis of various method also is spent.

Bogolubov has noticed that the generalization on his theory for a liquid
helium it is impossible, therefore the use of a phenomenological equations
of continuous medium it is necessary. This remark and to the subsequent
works concerns. The essential step in the theory of liquid helium by Feinman
was made.

\subsection{Feinman formula - the connection of the elementary excitations
energy and the formfactor}

Feinman (see in (Feinman, 1974) has established at $T=0$ the general
connection of the elementary excitation spectrum in liquid helium with the
static formfactor $S(p)$

\begin{equation}
\varepsilon (p)=\frac{p^{2}}{2mS(p)},\quad p=\hbar k.  \label{S.1.16}
\end{equation}
The static formfactor is defined via the spatial Fourier component of a
two-point correlation function or, that is equivalent, through the spatial
density of the number particles density fluctuations. For a liquid helium it
can be established on a basis of experiments on the scattering of x-ray or
neitrons.

In the paper (Bogolubov, Zubarev, 1955) the Feinman relation on the examle
of weakly nonideal Bose gas by the collective variable method was derivated.
Thus it is possible simultneously to receive expressions both for a
elementary exitation spectra, and for the formfactor.

\subsection{Kinetic derivation of the Feinman relation for the nonideal Bose
gas}

Being based on the kinetic equation for the Vigner function $f(r,p,t)$ with
the Langevin source the account of the particles number density fluctuations
it is possible to carry out At zero temperature ($T=0$) is received the
following expression for the required spectral density (Klimontovich and
Silin, 1952; Pins and Nozier, 1966; Nozier, Pines, 1980):

\begin{equation}
\left( \delta n\delta n\right) _{\omega ,k}=\frac{\hbar }{\nu (k)}\frac{%
Im\chi (\omega ,k)}{\left| \chi (\omega ,k)\right| ^{2}}  \label{S.2.22}
\end{equation}
Here $\chi (\omega ,k)$ is the dynamic susceptibility. By integration on
frequency in the region of a transparency we find the spatial spectral
density, which is connected with the static formfactor - the Feinman
relation for a weak nonideal Bose gas

\begin{equation}
\frac{\left( \delta n\delta n\right) _{k}}{n}\equiv S(k)=\frac{{}\hbar
^{2}k^{2}}{2m\hbar \omega (k)}.  \label{S.2.24}
\end{equation}

The Feinman relation was established under condition of $T=0$ and for this
reason it is non sufficient for the description of phase transition in
superfluid state, which occurs at non zero temperatures.

\subsection{Feinman relation in a classical limit}

Let's address to the Feinman relation. Let's rewrite it as

\begin{equation}
\omega (k)=\frac{\hbar ^{2}}{2m}\frac{\left( \delta n\delta n\right) _{k}}{n}%
.  \label{S.2.25}
\end{equation}
The transition in the right part to the classical limit is impossible. It is
connected to that in the formula (\ref{S.2.22}) only a zero oscillations
were taken into account. In the classical limit , instead of (\ref{S.2.22}),
we have the following expression (Klimontovich, Silin, 1960; Pines, Nozier,
1966):

\begin{equation}
\left( \delta n\delta n\right) _{\omega ,k}=\frac{2}{\omega \nu (k)}\frac{%
Im\chi (\omega ,k)}{\left| \chi (\omega ,k)\right| ^{2}}k_{B}T
\label{S.2.26}
\end{equation}
By integration on $\omega $, we shall receive expression for the connection
of the spectral density of fluctuations and the component Fourier for
potential of atoms interaction:

\begin{equation}
\frac{\left( \delta n\delta n\right) _{k}}{n}=\frac{1}{1+\frac{n\nu (\left|
k\right| )}{k_{B}T}}.  \label{S.2.29}
\end{equation}
Using the definition of the isothermal compressity coefficient $\beta _{T}$
we shall obtain at $k=0$ the following relations:

\begin{equation}
\frac{\left( \delta n\delta n\right) _{k=0}}{n}\equiv nk_{B}T\beta _{T}=%
\frac{1}{1+\frac{n\nu (0)}{k_{B}T}}.  \label{S.2.31}
\end{equation}
Let's compare this expression with the appropriate the Ornstain - Zernike
expression, containing a component Fourier of the ''direct correlation
function $C(k)$ (Klimontovich, 1982; Martynov, 1992)

\begin{equation}
\frac{\left( \delta n\delta n\right) _{k=0}}{n}\equiv nk_{B}T\beta _{T}=%
\frac{1}{1-C(k=0)}\equiv \frac{1}{1+\frac{n\nu _{eff}(0)}{k_{B}T}}.
\label{S.2.32}
\end{equation}
Here, a level with a direct correlation function $C(k=0),$ the definition
for the effective potential $\nu _{eff}(0)$ of interaction of atoms in a
liquid helium is given.

Last equalities for dense gases and liquids are used. For this purpose it is
necessary to have the additional equations for definition of a direct
correlation function $C(k=0)$, or the corresponding effective potential $\nu
_{eff}(0)$ (Klimontovich, 1982; Martynov, 1992).

For the critical region at the phase transitions the effective potential
with the critical temperature will be connected by the equality: $n\left|
\nu _{eff}(0)\right| =k_{B}T_{C}.$

From the condition of positivity of the function $\left( \delta n\delta
n\right) _{k=0},$ $nk_{B}T\beta _{T}$ follows the restrictions

\begin{equation}
C(k=0)<1,\quad \frac{n\left| \nu _{eff}(0)\right| }{k_{B}T}<1.
\label{S.2.33}
\end{equation}
At presence \ of the phase transition (it physical sense will be explained
below) at the approach to the critical point from the side of high
temperature the isothermal compressity $\beta _{T}$ on the Curi law grows.
At the critical region the functions $1-C(k=0),$ $1+n\nu _{eff}(0)/k_{B}T$
are proportional to the difference of temperature $T-T_{C}.$ From this
follow that for temperature $T<T_{C}$ the instability appears.

As well as in the chapters of the second volume, to the second order phase
transitions are devoted, the restriction of instability occurs for due to
nonlinearity. This is provided by the replacement:

\begin{equation}
\frac{T-T_{C}}{T_{C}}\rightarrow \frac{T-T_{C}}{T_{C}}+\frac{\left| \psi
\right| ^{2}}{n}.  \label{S.2.34}
\end{equation}
The unknown function $\left| \psi \right| ^{2}$ will be found below by the
solution of the corresponding kinetic equations.

In result for the critical region are received the following expressions \
for correlator $\left( \delta n\delta n\right) _{k=0}$ \ and the isothermal
compressibility coefficient $nk_{B}T\beta _{T}$: 
\begin{equation}
\frac{\left( \delta n\delta n\right) _{k=0}}{n}\equiv nk_{B}T\beta _{T}=%
\frac{1}{\frac{T-T_{C}}{T_{C}}+\frac{\left| \psi \right| ^{2}}{n}},\quad
\left| \psi \right| ^{2}=n_{S}.  \label{S.2.35}
\end{equation}
From stated follows that the prevailing role plays the attractive between
molecules.

We can remark, that on the existence \ of the second order phase transition
in a liquid helium shows an anomalous temperature dependence of heat
capacity - ''$\lambda $-curve'' (Keesom, 1949).

\subsection{Physical definition of constants in the Ginsburg - Landau (GL)
equation}

Follow to Ginsburg-Landau (GL) theory, we introduce the complex local
effective wave function:

\begin{equation}
\psi (R,t)=|\psi |e^{i\varphi },\quad |\psi |^{2}=\frac{N_{S}}{V}=n_{S}.
\label{S.1.29}
\end{equation}
$n_{S}$ - the density of number of particles in a condensate. In the
stationary state the effective wave function to the GL equation satisfies
(GL, 1950; Ginsburg, Sobyanin, 1976).

One of opportunities of the description of temporary evolution in the GL
theory - the transition to the reversible Hartree equation. To reveal the
physical sense of coefficients in the GL equation we can compare two ways of
account of the susceptibility: on the GL equation, and on the quantum
kinetic equation in the self-consistent approximation for a nonideal Bose
gas (Klimontovich and Silin, 1952). In result is received relations:

\begin{equation}
bn=\alpha _{L}=n\left| \nu _{eff}(0)\right| =k_{B}T_{C}.  \label{S.1.30}
\end{equation}
It specifies analogy of the phase transition in a liquid helium and the
phase transition in Van der Waals system (Klimontovich, 1998).

At transition in a superfluid state varies not only thermodynamics, but and
hydrodynamics - there are '' two-liquid flows '' normal and superfluid
components of helium. It puts a question on a physical nature of the
occurrence ''two-liquid'' state. For the answer it is necessary to use the
evolutionary equations with the account of dissipation. For the answer on it
the dissipative kinetic equations will be used.

The temporary dependence can be entered in the stationary GL equation by two
extreme ways.

\subsection{Nonlinear Schroedinger equation}

With $\alpha _{L}=k_{B}T_{C}$ we have the following nonlinear Schroedinger
equation - the Hartree equation:

\begin{equation}
i\hbar \frac{\partial \psi }{\partial t}=-\frac{\hbar ^{2}}{2m}\frac{%
\partial ^{2}\psi }{\partial R^{2}}+k_{B}T_{C}\left[ \frac{T-T_{C}}{T_{C}}+%
\frac{\left| \psi \right| ^{2}}{n}\right] \psi .  \label{S.1.2}
\end{equation}

\subsection{Relaxation G-L equation (RGLE)}

Let's carry out in the Hartree equation formal replacement $t$ by imaginary
time $it$. In result we come to the relaxation GL equation (RGLE): 
\begin{equation}
\frac{\partial \psi }{\partial t}=-\text{ }\gamma \left( \frac{T-T_{C}}{T_{C}%
}+\frac{\left| \psi \right| ^{2}}{n}{}{}\right) \psi +D\frac{%
{}{}{}{}{}{}{}{}{}{}\partial ^{2}\psi }{{}{}{}{}{}{}{}\partial R^{2}{}}{}.
\label{S.5.2}
\end{equation}
This is an example of the reaction diffusion equation. The coefficients $D$
and $\gamma $ are determined by the formulas:

\begin{equation}
D=\frac{\hbar }{2m},\qquad \gamma =\frac{\alpha _{L}}{%
{}{}{}{}{}{}{}{}{}{}{}{}{}{}{}{}\hbar },\quad \alpha _{L}=n\left| \nu
_{eff}(0)\right| =k_{B}T_{C}.  \label{S.5.3}
\end{equation}
Let's consider an opportunity of other derscription of temporary evolution
on a basis of the kinetic equation for the kinetic equation for the local
distribution function $f(n_{S},R,t)$:

\begin{equation}
\frac{\partial f}{\partial t}=2\frac{\partial }{\partial n_{S}}%
{}{}{}{}{}{}{}{}\left[ \gamma D_{n_{S}}n_{S}\frac{\partial f}{\partial n_{S}}%
{}{}{}{}{}\right] +{}{}\frac{{}{}{}{}{}{}{}{}{}\partial }{\partial n_{S}}%
\left[ 2\gamma \left( \frac{T-T_{C}}{T_{C}}+\frac{n_{S}}{n}\right) n_{S}f%
\right] +D{}\frac{{}{}{}{}\partial ^{2}f{}}{{}{}\partial R^{2}}.
\label{S.6.1}
\end{equation}
The diffusion coefficient $D_{n_{S}}=n/N_{ph}$ is defined by number of
particles in a point of continuous medium. In an equilibrium state:

\begin{equation}
f_{0}(n_{S})=C\exp \left( -\frac{h_{eff}}{D_{n_{S}}}\right) ,\qquad
h_{eff}=\gamma \left[ \frac{T-T_{C}}{T_{C}}n_{S}+\frac{n_{S}^{2}}{2n}\right]
.  \label{S.6.2}
\end{equation}
Here the designation for the effective Hamilton function in account on one
particle is entered. In the first moment approximation the distribution
function

\begin{equation}
f(n_{S},t)=\delta \left( n_{S}-\left\langle n_{S}\right\rangle _{R,t}\right)
.  \label{S.6.12}
\end{equation}
and we have the following equation for the first moment:

\begin{equation}
\frac{d\left\langle n_{S}\right\rangle }{dt}=2\left[ \gamma D_{n_{S}}-\gamma
\left( \frac{T-T_{C}}{T_{C}}+{}\frac{{}{}{}{}{}{}\left\langle
n_{S}\right\rangle }{n}{}{}{}{}{}{}{}{}{}{}\right) \left\langle
n_{S}\right\rangle \right] +D\frac{\partial ^{2}\left\langle
n_{S}\right\rangle _{R,t}}{{}{}{}\partial R^{2}}.  \label{S.6.13}
\end{equation}
It is an another example of the reaction diffusion equation. For stationary
and homogeneous state it is reduced to the algebraic equation for $%
\left\langle n_{S}\right\rangle $: 
\begin{equation}
D_{n_{S}}-\left( \frac{T-T_{C}}{T_{C}}+{}\frac{{}{}{}{}{}{}\left\langle
n_{S}\right\rangle }{n}{}{}{}{}{}{}{}{}{}{}\right) \left\langle
n_{S}\right\rangle =0,\qquad D_{n_{S}}=\frac{n}{N_{ph}},\quad \gamma =\frac{%
k_{B}T_{C}}{\hbar }.  \label{S.6.13a}
\end{equation}
Its solution allows to find average value of number of superfluid atoms $%
\left\langle n_{S}\right\rangle $ at all value of temperature. In
particular, in the critical point:

\begin{equation}
\left\langle n_{S}\right\rangle _{st}^{(2)}=\sqrt{nD_{n_{S}}}=n\sqrt{\frac{1%
}{N_{ph}}}\ll n.  \label{S.6.15}
\end{equation}
In the thermodynamic limit the average density of superfluid atoms is equal
to zero. Below the critical point:

\begin{equation}
\left\langle n_{S}\right\rangle _{st}^{(3)}=n\frac{T_{C}-T}{T_{C}}.
\label{S.6.16}
\end{equation}

\subsection{Distribution function of amplitudes and phases}

Instead the effective wave function we shall enter two real functions $%
X(R,t),Y(R.t)$:

\begin{equation}
\psi (R,t)=X(R,t)+iY(R.t),\qquad n_{S}(R.t)=X^{2}(R,t)+Y^{2}(R.t).
\label{O.1}
\end{equation}
The appropriate kinetic equation for the distribution function $f(X,Y,R,t)$
with is taking into account the spatial diffusion, has the following form
(see Klimontovich, 1995):

\[
\frac{\partial f}{\partial t}=\left\{ \frac{\partial }{\partial X}\left[ 
\frac{1}{2}\gamma D_{n_{S}}\frac{\partial f}{\partial X}\right] +\frac{%
\partial }{\partial Y}\left[ \frac{1}{2}\gamma D_{n_{S}}\frac{\partial f}{%
\partial Y}\right] \right\} + 
\]
\[
\left\{ \frac{\partial }{\partial X}\left[ \gamma \left( \frac{T-T_{C}}{T_{C}%
}+{}{}{}\frac{X^{2}+Y^{2})}{n}\right) Xf\right] +\frac{\partial }{\partial Y}%
\left[ \gamma \left( \frac{T-T_{C}}{T_{C}}+{}{}{}\frac{X^{2}+Y^{2})}{n}%
\right) Yf\right] \right\} + 
\]
\begin{equation}
D\frac{\partial ^{2}f}{\partial R^{2}},\quad D_{n_{S}}=\frac{n}{N_{ph}}%
,\quad \gamma =\frac{k_{B}T_{C}}{\hbar }.  \label{O.2}
\end{equation}
We can find now the equation for the distribution function of values $n_{S}$:

\begin{equation}
f(n_{S},R,t)=\int \delta (n_{S}-(X^{2}+Y^{2}))f(X,Y,R,t)dXdY  \label{O.5}
\end{equation}
In the self-consistent approximation for $n_{S}$ follows equation which
coincides with the equation (\ref{S.6.1}).

The relaxation processes below the critical point it is possible to divide
into two parts: on fast - for density of number of particles or amplitude,
and slow - for a phase or appropriate combination variable $X,$ $Y.$

For the description of the slow relaxation in the equation (\ref{O.2}) the
following replacement is introduced (see in Klimontovich, 1999):

\begin{equation}
\frac{T-T_{C}}{T_{C}}+{}{}{}\frac{X^{2}+Y^{2})}{n}\rightarrow \frac{T-T_{C}}{%
T_{C}}+\frac{n_{S}}{n}.  \label{0.6}
\end{equation}
In result is received the kinetic equation: 
\[
\frac{\partial f}{\partial t}=\left\{ \frac{\partial }{\partial X}\left[ 
\frac{1}{2}\gamma D_{n_{S}}\frac{\partial f}{\partial X}\right] +\frac{%
\partial }{\partial Y}\left[ \frac{1}{2}D_{n_{S}}\gamma \frac{\partial f}{%
\partial Y}\right] \right\} + 
\]
\begin{equation}
\left\{ \frac{\partial }{\partial X}\left( \gamma \frac{D_{n_{S}}}{%
\left\langle n_{S}\right\rangle _{st}}Xf\right) +\frac{\partial }{\partial Y}%
\left( \gamma \frac{D_{n_{S}}}{\left\langle n_{S}\right\rangle _{st}}%
Yf\right) \right\} +D\frac{\partial ^{2}f}{\partial R^{2}},\qquad D_{n_{S}}=%
\frac{n}{N_{ph}}{}{}{}{}{}{}{}{},\quad \gamma =\frac{k_{B}T_{C}}{\hbar }.
\label{O.8}
\end{equation}
The equilibrium decision of this equation - the Gauss distribution on two
variables:

\begin{equation}
f_{0}(X,Y)=C\exp \left( -\frac{X^{2}+Y^{2})}{\left\langle n_{S}\right\rangle
_{st}}\right) ,\qquad \int f_{0}(X,Y)dXdY=1.  \label{O.9}
\end{equation}
The average values $X,Y$ are equal to zero. The information on phase
transition contains in the expression for the dispersion $\left\langle
n_{S}\right\rangle _{st}$ . It at all values of temperature is defined by
the solution of the equations (\ref{S.6.13a}).

\section{Fast and slow fluctuations at phase transitions}

At phenomenological description of phase transitions the linear part of the
friction coefficient changes its sing.

In this respect there is an analogy with transition through the critical
point in Van der Waals system and with a second order phase transition in
segnetoelectrics We shall show that at transition via the critical point
changes radically and the character of relaxation and fluctuating processes.
The states at temperature $T>T_{C}$ will be characterized by fast processes.
At phase transition to states with temperature $T_{C}>T$ , on a level with
fast fluctuations appears and slow ones. We shall show that slow
fluctuations provide the coherence flow of helium in the Kapitza experiments.

\subsection{Relaxation of fast processes at phase transitions}

Let's returne to the evolutionary equation (\ref{S.6.13}) and consider a
small a deviation from the stationary and the spatial homogeneous solution $%
\left\langle n_{S}\right\rangle _{st}.$ The appropriate time relaxation and
half-width of a spectral line are defined by expressions:

\begin{equation}
\frac{1}{\tau _{n_{S}}}\left( k\right) =\Delta _{n_{S}}=2\gamma \left( \frac{%
T-T_{C}}{T_{C}}+{}2\frac{{}{}{}{}{}{}\left\langle n_{S}\right\rangle }{n}%
{}{}{}{}{}{}{}{}{}{}\right) +Dk^{2},\quad \quad \gamma =\frac{n\left| \nu
_{eff}(0)\right| }{\hbar }.  \label{O.14}
\end{equation}
At zero value of wave number the relaxation time grows at approach to the
critical point under the Curi law. In the critical point it has the maximal
finite value $\sqrt{N_{ph}}/2\gamma $.

The complex response $\left\langle n_{S}\right\rangle ^{(1)}$ on the
external action:

\begin{equation}
\chi _{(n_{S})}(\omega ,k)=\frac{1}{-i\omega +\Delta _{n_{S}}(k)},\quad
\Delta _{n_{S}}(k)=2\frac{k_{B}T_{C}}{\hbar }\left( \frac{T-T_{C}}{T_{C}}+{}2%
\frac{{}{}{}{}{}{}\left\langle n_{S}\right\rangle _{st}}{n}%
{}{}{}{}{}{}{}{}{}{}\right) +Dk^{2}.  \label{O.18a}
\end{equation}
We shall use connection of function $\chi _{(n_{S})}(0,0)$ and the
isothermal compressibility:

\begin{equation}
\chi _{(n_{S})}(0,0)=\frac{1}{\Delta _{n_{S}}(0)}=2\frac{k_{B}T_{C}}{\hbar }%
nk_{B}T_{C}\beta _{T}.\quad nk_{B}T_{C}\beta _{T}=\frac{1}{\frac{T-T_{C}}{%
T_{C}}+2\frac{{}{}{}{}{}{}\left\langle n_{S}\right\rangle _{st}}{n}{}{}}.
\label{O.18b}
\end{equation}
For the Landau region at approach to the critical point $\beta _{T}$ and the
relaxation time $\tau _{n_{S}}$ grow under the Curi law. In the critical
point $k=0$ the compressibility is finite

\begin{equation}
nk_{B}T_{C}\beta _{T_{C}}=\sqrt{N_{ph}}.  \label{O.18d}
\end{equation}

\subsection{Relaxation of slow processes}

From the kinetic equation (\ref{O.8}) we find the equations for average
values $\left\langle X\right\rangle _{R,t},$ $\left\langle Y\right\rangle
_{R,t}$ and with their help - the expressions for correlation time and the
width of spectral lines:

\begin{equation}
\frac{1}{\tau _{(X,Y)}}\left( k\right) =\Delta _{(X,Y)}\left( k\right)
=\gamma \left( \frac{D_{n_{S}}}{\left\langle n_{S}\right\rangle _{st}}%
{}{}{}{}{}{}{}{}{}{}\right) +Dk^{2},D_{n_{S}}=\frac{n}{N_{ph}}%
{}{}{}{}{}{}{}{}.  \label{O.19}
\end{equation}
The $\left\langle n_{S}\right\rangle _{st}$ \ - the solution of the equation
(\ref{S.6.13a}) at all temperatures. The expression for a complex
susceptibility has the form:

\begin{equation}
\chi _{(X)}(\omega ,k)=\chi _{(Y)}(\omega ,k)=\frac{1}{-i\omega +\gamma 
\frac{D_{n_{S}}}{\left\langle n_{S}\right\rangle _{st}}\left(
1+r_{C}^{2}k^{2}\right) },\qquad r_{C}^{2}=D\frac{\left\langle
n_{S}\right\rangle _{st}}{\gamma D_{n_{S}}}.  \label{O.19a}
\end{equation}
The designation for a square of the correlation radius here is entered. The
isothermal comressibility is defined now by expression:

\begin{equation}
nk_{B}T_{C}\beta _{T}=\frac{\left\langle n_{S}\right\rangle _{st}}{D_{n_{S}}}%
{}{}{}{}{}{}{}{}{}{}{}{}{}{}{}{}{}{}{}{}.  \label{O.19c}
\end{equation}
For the region of applicability of the Landau theory at approach to the
critical point from the side of high temperatures the isothermal
compressibility grows under the Curi law. In the critical point it is finite
also and is defined by expression:

\begin{equation}
nk_{B}T_{C}\beta _{T_{C}}=\frac{k_{B}T_{?}}{n\left| \nu _{eff}(0)\right| }%
\sqrt{N_{ph}},\quad T=T_{C}.  \label{O.19d}
\end{equation}
However, for the region of temperatures, below the critical one

\begin{equation}
nk_{B}T_{C}\beta _{T}=N_{ph}\frac{T_{C}-T}{T_{C}},\qquad T<T_{C},
\label{O.19e}
\end{equation}
the isothermal compressibility continues grow at a measure of downturn of
temperature.

From received formulas follows that at temperature enough below critical one
the ration of fast and slow correlation times is proportional to $1/N_{ph}.$

\subsection{Fast and slow fluctuations}

The task of an account of fluctuations at phase transition in a superfluid
state is similar, solved in Ch.24 (Klimontovich, 1999) '' Kinetic
fluctuations in the critical point ''.

\subsubsection{Fast fluctuations}

The account is similar, carried out in the section (19.6) (Klimontovich,
1999).

We enter in the equation (\ref{S.6.13}) for the function $\left\langle
n_{S}\right\rangle _{R,t}$ the Langevin source, which reflects the atomic
structure of a liquid helium. The relaxation time of the parameter of order
coincides with expression (\ref{O.14}).

The complex response to a random source is defined above mentioned,
expressions at all values of temperature. The quantity $\left\langle
n_{S}\right\rangle $ is defined by the solution of the equations (\ref
{S.6.13a}). At sufficient distance from the critical point the response $%
\chi _{(n_{S})}(\omega =0,k=0)$ varies under the Curi law. In the critical
point the isothermal comressibility has finite value (\ref{O.18d}). A square
of correlation radius at approach to the critical point in the region field
of the Landau theory varies under the Curi law, but it is finite in the
critical point.

\subsubsection{Slow fluctuations}

Let's show, that the existence of superfluidity is possible due to slow
fluctuations.

For this let's address to the kinetic equation (\ref{O.8}). Let's enter into
it the Langevin source, which intensity is determined by two dissipative
characteristics ('' by integrals of collisions ''). In self-consistent
approximation the equation for the first moments - functions $X(R,t),$ $%
Y(R,t)$ have the following form:

\begin{equation}
\frac{\partial X}{\partial t}+\gamma \frac{D_{n_{S}}}{\left\langle
n_{S}\right\rangle _{st}}X=D\frac{\partial ^{2}X}{\partial R^{2}}%
+y_{(X)}(R,T),\quad \frac{\partial Y}{\partial t}+\gamma \frac{D_{n_{S}}}{%
\left\langle n_{S}\right\rangle _{st}}Y=D\frac{\partial ^{2}Y}{\partial R^{2}%
}+y_{(Y)}(R,T).  \label{10.28}
\end{equation}
The width of spectral lines and the dynamic susceptibility are defined by
former expressions (\ref{O.19}) - (\ref{O.19a}).

For the region of the Landau theory at approach to the critical point from
the side of high temperatures the square of correlation radius grows on the
Curi law:

\begin{equation}
r_{C}^{2}=\frac{D}{\gamma }\frac{T_{C}}{T-T_{C}},\qquad T>T_{C}.
\label{O.33}
\end{equation}
In the critical point the square of radius of correlation is finite:

\begin{equation}
r_{C}^{2}=\sqrt{N_{ph}}\frac{D}{\gamma },\quad T=T_{C}.  \label{O.34}
\end{equation}
At downturn of temperature from critical

\begin{equation}
r_{C}^{2}=N_{ph}\frac{D}{\gamma }\frac{T_{C}-T}{T_{C}}\sim N_{ph}\lambda
_{S}^{2}\frac{T_{C}-T}{T_{C}},\quad T<T_{C}.  \label{O.35}
\end{equation}
The following estimation of parameter $D/\gamma $ here is used:

\begin{equation}
\frac{D}{\gamma }\sim \frac{\hbar ^{2}}{mk_{B}T_{C}}\sim \lambda _{S}^{2}.
\label{O.36}
\end{equation}
$v_{S}$ - speed of the sound, $\lambda _{S}$ is the appropriate length of de
Broglie.

Thus, value of the isothermal compressibility and the square of correlation
radius for slow fluctuations at temperatures lower of the critical
temperature ($T<T_{C}$) are defined by number of atoms in a point of
continuous medium $N_{ph}$. Thus $r_{C}^{2}$ at $T<T_{C}$ is macroscopic
characteristic.

Now we can give an estimation of $N_{ph}$ for the hydrodynamic description.

\subsection{Physically infinitesimal scales}

As the description of superfluidity is carried out on hydrodynamical level,
it is necessary to use and appropriate definition of physically
infinitesimal scales (Klimontovich, 1982,1995, 1999). Thus the required
scales depend from external parameter of length $L$ - one of the
characteristic sizes of a vessel with a liquid helium.

According (Klimontovich, 1995) the formula for $r_{C}^{2}$ at low
temperatures it is possible to rewrite as:

\begin{equation}
r_{C}^{2}=\left( nL^{3}\right) ^{2/.5}\lambda _{S}^{2}\frac{T_{C}-T}{T_{C}}%
\sim \left( nL^{3}\right) ^{2/.5} \lambda _{S}^{2},\quad T<T_{C}.
\label{S.6.21}
\end{equation}
To ensure the spatial coherence of the superfluid flow on capillary of a
diameter $d $, it is necessary to limit size $d $ to a condition:

\begin{equation}
d<r_{C}=\left( nL^{3}\right) ^{1/5}\lambda _{S}.  \label{S.6.22}
\end{equation}

This inequality is carried out for the Kapitza experiments

\subsection{Spectral density of slow fluctuations}

The account is similar carried out in Chs.19, 24 (Klimontovich, 1999),
devoted the kinetic theory of fluctuations at phase transitions in
segnetoelectrics and in the Van der Waals system. The structure of Langevin
sources in the equations for the first moments:

\begin{equation}
\text{ }\left( y_{(X,Y)}y_{(X,Y)}\right) _{\omega ,k}=2\Delta _{(X,Y)}\left(
k\right) \frac{1}{2}\frac{\left\langle n_{S}\right\rangle _{st}}{n}.
\label{S.6.24}
\end{equation}
The appropriate expressions for spectral density of slow fluctuations look
like:

\[
\left( \delta X\delta X\right) _{\omega ,k}=\frac{2\Delta _{(X)}\left(
k\right) }{\omega ^{2}+\Delta _{(X)}^{2}\left( k\right) }\frac{\left\langle
\left( \delta X\right) ^{2}\right\rangle }{n},\quad \left\langle \left(
\delta X\right) ^{2}\right\rangle =\frac{1}{2}\left\langle
n_{S}\right\rangle _{st}. 
\]
\begin{equation}  \label{S.6.25}
\end{equation}

\[
\left( \delta Y\delta Y\right) _{\omega ,k}=\frac{2\Delta _{(Y)}\left(
k\right) }{\omega ^{2}+\Delta _{(Y)}^{2}\left( k\right) }\frac{\left\langle
\left( \delta Y\right) ^{2}\right\rangle }{n},\qquad \left\langle \left(
\delta Y\right) ^{2}\right\rangle =\frac{1}{2}\left\langle
n_{S}\right\rangle _{st}. 
\]
The spatial spectral density does not depend from wave numbers - " spatial
white noise ", therefore the spatial correlators:

\begin{equation}
\left\langle \delta X\delta X\right\rangle _{R-R^{\prime }}=\left\langle
\delta Y\delta Y\right\rangle _{R-R^{\prime }}=\frac{1}{2}\frac{\left\langle
n_{S}\right\rangle _{st}}{n}\delta (R-R^{\prime }).  \label{S.6.28}
\end{equation}
Thus, the spatial correlations are different from zero only in limits of a
point of continuous medium (in physically infinitesimal volume $V_{ph} $),
as function $\delta (R-R ^ {\prime}) \mid _ {R = R ^{\prime}} = V_{ph}^{-1}. 
$ In result for the one-dot correlator slow (large-scale) fluctuation $%
\left\langle \delta X\delta X\right\rangle_{R = R ^{\prime}} $ we have the
following expression:

\begin{equation}
\left\langle \delta X\delta X\right\rangle _{R=R^{\prime }}=\left\langle
\delta Y\delta Y\right\rangle _{R=R^{\prime }}=\frac{1}{2}\frac{\left\langle
n_{S}\right\rangle }{N_{ph}},\quad N_{ph}=nV_{ph\text{{}}}.  \label{S.6.29}
\end{equation}
The dispersion of fluctuations, smoothed on volume of point of continuous
medium, in $N _ {ph} $ of times there is less fluctuations of one-dot
distribution - the Boltzmann distribution.

\subsection{Intermediate conclusion}

Was shown, that at temperatures below critical in a liquid helium are
available two fluctuation process - ''fast'' and ''slow''. Behavior of fast
fluctuations is similar what is offered by the theory of phase transitions,
advanced by Landau. The thermodynamic and the fluctuation characteristic of
fast processes demonstrate anomalous dependences on temperature - the Curi
law for the isothermal compressibility, the static susceptibility and the
square of radius of correlation.

The basic difference in behavior of fast fluctuations from the prediction of
the Landau theory and the more general  '' the fluctuation theory of phase
transitions ''  (Patashinskii, Pokrovskii, 1982) consists in absence of ''
the problems of infinity ''. Just in difference from the traditional theory
here the isothermal cpmpressibility, the static susceptibility and the
square of radius of correlation have finite value in the critical point!

More over, in the Landau theory at removal from the a critical point (in the
region of high temperatures and in the region of low temperatures) the
values of these characteristics decrease under the Curi up to their
''normal'' values . In a ''normal'' state the correlation radius of is the
order of the average distance between atoms. On this reason fast
correlations can not ensure the spatial and temporary coherence of more
ordered asymmetrical state, which exists at temperatures $T<T_{C}$.

On the contrary, behavior of slow fluctuations at temperatures $T<T_{C}$ \
is essentially other. In process of removal downwards from the critical
temperature the value of correlation radius continues to increase and
aspires to macroscopic value. In a thermodynamic limit this value, formally,
is equal to infinity. However, such limiting transition contradict to the
model of continuous medium. It is enough, that radius of correlation was
macroscopic, in particular, would exceed a diameter of a capillary in the
Kapitza installations.

The basic features of the phenomenon of superfluidity are shown not in the
thermodynamics only, but also in hydrodynamics. It is emphasized also by
term, which has entered by Kapitza - ''superfluidity''. So, the basic task
of the theory is reduced here to two questions:

1.Why the viscousless flow can exists a dissipative medium?

2.What physical distinction of superfluid and normal component in
hydrodynamics of helium?

Let's try to give the answer to these basic questions.

\section{The viscousless flow in the viscous medium}

\subsection{Dynamic-reaction-diffusion equation in the theory of
Superfluidity}

Above it was supposed, that the distribution of helium is the spatially
homogeneous. However, a liquid helium is viscous medium , therefore at the
spatially non-uniform state the viscous enters in the game. It is necessary
to show, that the carried out (at $k=0)$ the division on '' normal '' and ''
superfluid '' components at the account of the viscous is valid.

In liquid helium the wave length of de Broglie $\lambda _{B}^{(T)}$ is of
the order of the average distance between atoms $r_{av}$. However, in
continuous medium both these sizes it is a lot of less of physically
infinitesimal scale $l_{ph},$ as in a point of continuous medium there are
many atoms. On this reason the contributions determined by the Planck
constant, in many cases are not essential. The appropriate dissipative
kinetic equation for local distribution functions $f(X,Y,R,v,t)$ of values
of real and imaginary parts of the effective wave function has the following
structure:

\[
\frac{\partial f}{\partial t}+V\frac{\partial f}{\partial R}+F(R,t)\frac{%
\partial f}{\partial v}{}=I_{(v)}+\left\{ \frac{\partial } {\partial X}\left[
\frac{1}{2}\gamma D_{n_{S}}\frac{\partial f}{\partial X}\right] +\frac{%
\partial }{\partial Y}\left[ \frac{1}{2}D_{n_{S}}\gamma \frac{\partial f}{%
\partial Y}\right] \right\} + 
\]
\begin{equation}
\left\{ \frac{\partial }{\partial X}\left( \gamma \frac{D_{n_{S}}}{%
\left\langle n_{S}\right\rangle _{st}}Xf\right) + \frac{\partial }{\partial Y%
}\left( \gamma \frac{D_{n_{S}}}{\left\langle n_{S}\right\rangle _{st}}%
Yf\right) \right\} +D\frac{\partial ^{2}f}{\partial R^{2}}.  \label{T.3}
\end{equation}
The diffusion and friction coefficient are defined by former expressions:

\begin{equation}
D=\frac{\hbar }{2m},\quad D_{n_{S}}=\frac{n}{N_{ph}},\quad \gamma =k_{B}T.
\label{T.4}
\end{equation}

\subsection{Transition to the hydrodynamical equations}

The equation of a continuity now has the following form:

\begin{equation}
\frac{\partial }{\partial t}\left\langle n_{S}\right\rangle _{R,t}+\frac{%
\partial }{\partial R}j_{S}(R,t)=2\left[ D_{n_{S}}-\gamma \left( \frac{%
T-T_{C}}{T_{C}}+\frac{\left\langle n_{S}\right\rangle _{R,t}}{n}\right)
\left\langle n_{S}\right\rangle _{R,t}\right] .  \label{T.7}
\end{equation}
The designation for a flow of matter with the account as the convective, so
and spatial diffusion flows, here is used:

\begin{equation}
j_{S}(R,t)=\left\langle n_{S}\right\rangle _{R,t}u_{S}(R,t)-D\frac{\partial
\left\langle n_{S}\right\rangle _{R,t}}{\partial R}.  \label{T.8}
\end{equation}
The right part of the equation of a continuity describes '' the chemical
reaction '' - a birth and a destruction of average density of the superfluid
component. In the stationary state the equation of a continuity can be
reduced to the two equations. The first :

\begin{equation}
\frac{{}\partial u_{S}(R,t)}{{}\partial R}=0.  \label{T.11}
\end{equation}
allows to express the quantity of a velocity in capillary through difference
of helium density on the ends of a capillary.

Let's take into account, that the relaxation of the function $\left\langle
n_{S}\right\rangle _{R,t}$ is fast. In the stationsry state $\left\langle
n_{S}\right\rangle _{t}$ for all values of temperature by the solution of
the equation (\ref{S.6.13a}) can be defined.

In result the Navier-Stokes equation for the velocity of the superfluid
component is received:

\begin{equation}
\frac{\partial u_{S}}{\partial t}+\left( u_{S}\frac{\partial }{\partial R}%
\right) u_{S}=\frac{F}{m}+\nu \frac{\partial ^{2}u_{Si}}{\partial R^{2}}%
,\qquad j_{S}(R,t)=\left\langle n_{S}\right\rangle _{t}u_{S}(R,t).
\label{T.12}
\end{equation}
The equation of motion is not include the dependence on temperature. The
flow of superfluid helium through density $\left\langle n_{S}\right\rangle
_{t}$ depends on temperature. Here $D\rightarrow \nu .$

At values of speed $u_{S}$ much smaller of the critical one $u_{C}$, it is
possible to neglect by the nonlinear terms. In result at $F=0$ we come for
the velocity $u_{S}$ to the diffusion equation:

\begin{equation}
\frac{\partial u_{S}}{\partial t}=\nu \frac{\partial ^{2}u_{S}}{\partial
R^{2}}\qquad j_{S}(R,t)=\left\langle n_{S}\right\rangle _{R,t}u_{S}(R,t).
\label{T.13}
\end{equation}
Thus the quetion arises: Why the viscousless flow can exists a viscous
medium?

\section{Flicker noise and Viscousless flow in viscous medium}

\subsection{Flicker noise}

Let's result only minimum of the necessary information from the theory
Flicker noise \ (Klimontovich, 1982; Kogan, 1985).

Flicker noise from the side of high frequencies is limited by diffusion time 
$\tau _{\nu }=L^{2}/\nu ,$ $L-$ a minimal characteristic scale of a sample.
Here it will be a diameter of a capillary $d$. The region of the existence
of the flicker noise on frequencies is defined by inequalities:

\begin{equation}
\frac{1}{\tau _{life}}\ll \omega \ll \frac{1}{\tau _{\nu }}=\frac{\nu }{d^{2}%
}.  \label{T.14}
\end{equation}
$\tau _{life}$ - time of life of installation. In the region of flicker
noise there is a new scale and appropriate volume:

\begin{equation}
L_{\omega }=\sqrt{\frac{\nu }{\omega }}\gg L,\qquad V_{\omega }=L_{\omega
}^{3}\gg V.  \label{T.15}
\end{equation}
Let's result the appropriate chain of inequalities for volumes:

\begin{equation}
V\ll V_{\omega }\ll V_{life}.  \label{T.16}
\end{equation}
The equilibrium (natural) flicker noise arises at the diffusion in limited
volume. Here it the diffusion of velocity $u_{S}$.

The expression for the spatial - temporary spectrum fluvtuations of velocity 
$U_{S}$ follows from the appropriate Langevin equation and has the form
(Klimontovich, 1982,1990):

\begin{equation}
\left( \delta u_{S}\delta u_{S}\right) _{\omega ,k}=\frac{\left( yy\right)
_{\omega ,k}}{\omega ^{2}+\left( \nu k^{2}\right) ^{2}},\quad \left(
yy\right) _{\omega ,k}=2\nu k^{2}AV_{\omega }\left\langle \delta u_{S}\delta
u_{S}\right\rangle _{V}\exp \left( -\frac{\nu k^{2}}{2\omega }\right) .
\label{T.17}
\end{equation}
Here $\left\langle \delta u_{S}\delta u_{S}\right\rangle _{V}$ - the
correlator fluctuations, average on volume of a sample. The $A$ will be
determined below from the normalization condition. From the last formula
follows, that has a place strong dependence on frequency and on wave number.
Thus dispersion of wave numbers is proportional to frequency $\omega $

\begin{equation}
\left\langle \left( \delta k\right) ^{2}\right\rangle \sim \frac{1}{%
L_{\omega }^{2}}=\frac{\omega }{\nu }.  \label{T.19}
\end{equation}
Thus in the region of flicker noise has a place '' the original Bose
condensation ''. This shows the presence of the spatial coherence.

In expression for spatially temporary spectral density

\begin{equation}
\left( \delta u_{S}\delta u_{S}\right) _{\omega ,k}=\frac{2\nu k^{2}}{\omega
^{2}+\left( \nu k^{2}\right) ^{2}}AV_{\omega }\left\langle \delta
u_{S}\delta u_{S}\right\rangle _{V}\exp \left( -\frac{\nu k^{2}}{2\omega }%
\right) .  \label{T.20}
\end{equation}
it is possible to execute the integration on $k.$ In result is received
expression for the appropriate temporary spectral density:

\begin{equation}
\left( \delta u_{S}\delta u_{S}\right) _{\omega }=\frac{\pi \left\langle
\delta u_{S}\delta u_{S}\right\rangle _{V}}{\ln \left( \tau _{life}/\tau
_{\nu }\right) }{}\frac{1}{\omega },\qquad \frac{1}{\tau _{life}}\ll \omega
\ll \frac{1}{\tau _{\nu }}.  \label{T.21}
\end{equation}
$A$ is determined from the normalization condition:

\begin{equation}
\int_{1/\tau _{life}}^{1/\tau _{\nu }}\left( \delta u_{S}\delta u_{S}\right)
_{\omega }\frac{d\omega }{\pi }=\left\langle \delta u_{S}\delta
u_{S}\right\rangle _{V}.  \label{T.22}
\end{equation}
It is supposed, thus, that the basic contribution to the correlator $%
\left\langle \delta u _ {S} \delta u _ {S} \right\rangle _ {V} $ has to the
region of flicker noise.

\subsection{Temporary correlation}

The temporary correlation is connected to temporary spectral density by
relation:

\begin{equation}
\left\langle \delta u_{S}\delta u_{S}\right\rangle _{\tau }=\int_{1/\tau
_{life}}^{1/\tau _{\nu }}\left( \delta u_{S}\delta u_{S}\right) _{\omega }%
\frac{d\omega }{\pi }.  \label{T.23}
\end{equation}
From here follows that 
\[
\left\langle \delta u_{S}\delta u_{S}\right\rangle _{\tau }=\left( C-\frac{%
\ln \left( \tau /\tau _{\nu }\right) }{\ln \left( \tau _{life}/\tau _{\nu
}\right) }\right) \left\langle \delta u_{S}\delta u_{S}\right\rangle
_{V},\quad \text{\quad\ }\tau _{\nu }\ll \tau \ll \tau _{life}, 
\]
\begin{equation}  \label{T.24}
\end{equation}
\[
C=1-\frac{\gamma }{\ln \left( \tau _{life}/\tau _{\nu }\right) },\quad \text{
}\gamma =0.577. 
\]
Here are used the Euler constant.

In the region of flicker noise the dependence from $\tau $ is logarithmic at
the large value of argument. It gives the basis to speak about presence of
the residual correlations.

The characteristic time of the correlations is defined by expression:

\begin{equation}
\tau _{cor}=\int_{\tau _{D}}^{\tau _{life}}\left\langle \delta u_{S}\delta
u_{S}\right\rangle _{\tau }d\tau /\left\langle \delta u_{S}\delta
u_{S}\right\rangle _{\tau }.  \label{T.25}
\end{equation}
In result we find, that

\begin{equation}
\tau _{cor}\sim \tau _{life}/\ln \frac{\tau _{life}}{\tau _{\nu }}.
\label{T.27}
\end{equation}
The time of correlation at the unlimited time $\tau _{life}$ tends to
infinity.

The stated in the present section shows, that in the region of flicker noise
has a place as the spatial, and also the temporary coherence. It also gives
the basis for to establish connection of two coherent phenomena: flicker
noise and superfluidity.

\subsection{Flicker noise and superfluidity}

Let's return to the equation (\ref{T.13}) for velocity of the speed
superfluid helium. The account of fluvctuations of velocity in the region of
the flicker noise leads to the formula (\ref{T.21}) for the temporary
spectrum fluctuations of the superfluid components of velocity of liquid
helium.

For the dissipative term we shall use the simplest approximation of kind $%
"1/\tau _{rel}"$. In result for the velocity of superfluid component is
received the following relaxation equation:

\begin{equation}
\frac{\partial u_{S}}{\partial t}=-\text{ }\frac{1}{\tau _{rel}}u_{S}.
\label{T.28}
\end{equation}
In zero approximation on dimensionless parameter

\begin{equation}
\frac{\tau _{obs}}{\tau _{life}},\quad \text{ \quad }\tau _{life}\gg \tau
_{obs}\gg \tau _{D}  \label{T.29}
\end{equation}
in the equation (\ref{T.28}) it is possible to neglect by dissipation. Thus
we come, to the equation

\begin{equation}
\frac{\partial u_{S}}{\partial t}=0,\qquad u_{S}=const.  \label{T.30}
\end{equation}
The value of constant velocity is defined by boundary conditions on the ends
of capillary.

Let's estimate the diffusion time $\tau _{\nu }.$ It defines the high
boundary of the flicker noise region: $\omega _{\max }\sim 1/\tau _{\nu }.$
\ The diameter of a capillary in the Kapitza experiences of the order 10 $^{%
\text{-4}}$ -10 $^{\text{-5}cm}$. The diffusion coefficient can be estimated
on one of two formulas: $D\approx \hbar /2m^{\ast };$ $v_{T}l$. ($l$%
-effective length of free paths.) Thus the list value of the time
observation of superfluidity 
\begin{equation}
\left( \tau _{obs}\right) _{\min }\geq \tau _{\nu }=d^{2}/\nu \sim {10}^{-5}-%
{10}^{-6}sec.  \label{12.13.37}
\end{equation}

\subsection{Superfluidity is viscousless flow in viscous medium}

Let's address to definition of concept ''superfluidity'' - viscousless flow
in viscous medium.

One of the definitions of this concept is ''the measuring''. It is connected
with the observation time $\tau _{obs}$.

However, as in process of increase of the observation time it fails to find
out the reduction of velocity, it is natural to assume, that the constancy
of velocity has a place within the limits of greatest temporary interval $%
\tau _{life},$ i.e. '' of the time life of installation

\section{Conclusion}

The results can be presented as two parts:

{\bf 1. Thermodynamics of helium-II.}

On the basis of the corresponding kinetic equations was shown, that for the
spatially homogeneous state ($k=0$), when the diffusion processes drop out,
at temperatures $T<T_{C}$ there are two kinds relaxation and fluctuation
processes: the fast and the slow.

For the fast processes the radius of correlation and the isothermal
compressibility at approach to a critical point grow under by the Curi law,
but in the critical point they have finite values! At removal from the
critical point ''downwards'' the radius of correlation again becomes about
average distances between atoms. For this reason the fast fluctuations can
not to ensure the coherence of an asymmetrical phase on macroscopic scales.

On the contrary, the slow fluctuations at $T<T_{C}$ become macroscopic. The
appropriate radius of correlation are more than a diameter of capillary. Is
provided, thus, spatial coherence of superfluid components in the Kapitza
experiences.

{\bf 2. Hydrodynamics of helium -II.}

Was shown, that the existence of superfluidity in viscous medium is possible
due to occurence of the natural fliccker noise. Thus there is ''the
Bose-condansation'' in the space of wave vectors. The dispersion of
distribution on wave numbers is proportional to frequency. In results in the
equations for the Fourier component of velocity it is possible the following
replacement of hydrodynamical friction: $\nu k^{2}\rightarrow \omega .$
Accordingly to this arise ''residual'' temporary correlations are limited on
duration by time life of installation $\tau _{life}$ only. This time defines
and the relaxation time for the velocity of superfluid component.

The new opportunity of physical treatment of two-liquid model of Tisza and
Landau is opened. The independent existence of two motions it is possible
only in linear approximation. The superfluid flow is broken, when it begins
the change vortical motion of normal component. The appropriate critical
velocity: $u_{C}\sim \hbar /md.$ It corresponds to known estimation of
critical speed (see in Lifshitz and Pitaevskii, 1978).

The physical treatment, submitted in these chapters, is possible to hope,
allows better to understand the essence of the phenomenon of superfluidity,
opened by Kapitza and detailed investigated by him in a number of remarkable
papers.

\section{References}

1.Andronikashvily E.L. (JETP, 18 (1940) 424).

2.Belyaev S.T. (JETP 34 (1958) 417).

3. Bogolubov N.N. To theory of superfluidity. (Izv. AN, FIZ. v.11 (1047) 77).

4. Bogolubov N.N., Zubarev D.N. (JETP 28 (1955) 129).

5 Ginsburg V.L., Landau L.D. (JETF 20 (1950) 106).

6. Ginsburg V.L., Sobyanin A.A. ( Uspehi Fiz. Nauk 120 (1976) 153).

7. Einstein A. Quantentheorie der einatomogen idealen Gases.(Berl.Ber. 22
(1924) 261; 23 (1925) 3, 18).

8. Kapitza P.L. (DAN U.S.S.R. 18 (1938) 21; Nature 141 (1938) 74).

9. Kapitza P.L. (JETP 11 (1941) 1).

10. Kapitza P.L. (JETP 11 (1941) 581).

11. Kapitza P.L. (JETF 20 (1944) 2).

12. Keesom V. Helium ("International Literature", Moscow, 1949).

13.Kirpatrick T.R. and Dorfman J.R. (J. Low Temp. Phys. 58 (1955) 301; 58.
(1955) 399; 59 (1955) 1).

14 Klimontovich Yu.L., Silin V.P. (JETP 23 (1952) 151; Uspehi Fiz. Nauk 70
(1960) 247).

15. Klimontovich Yu.L. Statistical Physics (''Nauka'' , Moscow, 1982;
Harwood, New York, 1986).

16. Klimontovich Y.L. (Pis'ma v JTP 9 (1983) 406; Sov. Techn. Phys. Let. 9
(1983) 174).

17. Klimontovich Y.L. (Pis'ma v JETP 51 (1990) 43; Physica A 167 (1990) 782).

18. Klimontovich Yu.L. Turbulent Motion and the Structure of chaos.
(''Nauka'', Moscow, 1990; Kluwer, Dordrecht, 1991).

19. Klimontovich Yu.L. (TMP 115 (1998) 437).

20. Klimontovich Yu.L. Statistical Theory of Open Systems. (V.I ''Yanus''
Moscow, 1995; Kluwer, Dordrecht, 1995); V.II ''Yanus'', Moscow, 1999; V.III,
2001).

21. Kogan Sh.M. (Uspehi Fiz. Nauk 145 (1985) 286).

22. Landau L.D. (JETP 11 (1941), 592).

23. Landau L.D. (JETP 14 (1944) 112).

24. Lifshitz E.M., Pitaevskii L.P. Statistical Physics. Part 2 (''Nauka''
Moscow, 1978).

25. London F. Superfluids, Vol.I Superconductivity. (Wiley New York, 1954).

26. Martynov G.A. Fundamental Theory of Liquids. (Adam Hilger, Bristol, New
York, 1992).

27. Nozieres Ph., Pines D. Theory of quantum liquids. V.II. Superfluid Bose
Liquids (Addison New York, 1980).

28. Patashinskii A.Z., Pokrovskii V.L. Fluctuation Theory of Phase
Transitions. (''Nauka'' Moscow, 1982).

29. Pines D. and Nosier P. The Theory of Quantum Liquids (Benjamin, New
York, 1966).

301. Ticza L. Nature 171 (1938) 913; Journ.Phys.Rad (8) 1 (1940) 913).

31. Feinman P. Statistical Mechanics (''Mir'' Moscow, 1974).

32. Tserkovnikov Y.A. (TMP 93 (1992) 412; TMP 105 (1955) 77).

\bigskip

\end{document}